\documentclass[prb,twocolumn]{revtex4-1}

\usepackage{graphicx}% Include figure files
\usepackage{dcolumn}% Align table columns on decimal point
\usepackage{bm}% bold math
\usepackage{wrapfig}
\usepackage{amsmath,amsfonts,amssymb,bm,ulem}
\usepackage[usenames]{color}
\usepackage{bm}
\usepackage{slashed}
\usepackage{mathrsfs}
\usepackage{dcolumn}% Align table columns on decimal point
\usepackage{bm}% bold math
\usepackage{dsfont}
\usepackage[caption = false]{subfig}
\usepackage[dvipsnames]{xcolor}

\usepackage{braket}

\begin{document}

\title{Coherent dynamics in a dual-coupling spin-boson model}

\author{Leonard Ruocco}
 \affiliation{Department of Physics and Astronomy, University of British Columbia, Vancouver B.C. V6T 1Z1, Canada and
Stewart Blusson Quantum Matter Institute, University of British Columbia, Vancouver B.C. V6T 1Z4, Canada*}

\date{\today}

\begin{abstract}
\noindent
We study the dynamics of a particle tunneling between the ground states of a symmetric double potential well system, in the presence of simultaneous diagonal and non-diagonal couplings to an Ohmic oscillator bath. We use the noninteracting-blip approximation to investigate coherence effects across a wide range of system-environment coupling strengths at physiological temperatures. We show how the presence of a non-diagonal coupling mechanism significantly alters the dynamics of the tunneling particle, producing coherent oscillations despite strong thermal fluctuations in the environment. Fluctuations in both the particle's polarization, as well as tunneling energies, lead to competing influences on the coherent nature of tunneling particle dynamics, with non-diagonal couplings introducing a relatively long-lived oscillatory mode in an otherwise incoherent setting. 
\end{abstract}
\maketitle
\noindent
\textit{Introduction}: Understanding and controlling open quantum systems is currently a key area of research spanning the scientific domain from quantum computing,\cite{nielsen10} to molecular physics and emergent quantum phenomena in biology.\cite{may04} A central challenge within these fields is understanding the mechanisms driving decoherence---the process by which a quantum system becomes entangled with its environment and loses its quantum information.\cite{schlosshauer19}

Accurately modeling the coherent dynamics of a quantum system subject to complex environmental interactions remains a significant challenge. One of the most powerful analytical models used to study quantum systems under the influence of a large number of environmental modes is the much celebrated spin-boson (SB) model.\cite{leggett87} The SB model describes a particle tunneling between the ground states of two potential wells coupled to a bath of harmonic oscillators. The model, and its corresponding solution using the noninteracting-blip approximation (NIBA), permits the exploration of both coherent and incoherent dynamics in a central two-level quantum system (TLS) of interest, such as a qubit\cite{shnirman02} or an impurity spin\cite{lehur18} coupled with arbitrary strength to an environment simulating classical noise on the system. It therefore offers a tool with which to study coherent dynamics in a wide class of quantum systems subject to significant environmental noise, without invoking limiting assumptions such as the Markov approximation.\cite{breuer02}

One significant limitation of the SB model however, is its restriction to only a `diagonal' system-environment coupling. This form of coupling modulates the on-site energy terms of the tunneling particle due to a fluctuating polarization field, and are otherwise known as Holstein couplings\cite{holstein59b} in the context of polaron physics. Another significant form of system-environment interaction, which is absent in the SB model, is the Peierls coupling\cite{peierls55} which modulates the tunneling term of the central two-state system. These types of interactions are otherwise known as non-diagonal (or non-local) couplings and it has become increasingly evident in recent years that the inclusion of such terms changes the physics substantially. Non-diagonal couplings, which describe fluctuations in the tunneling matrix element, are proportional to the overlap of the ground state wavefunctions in the two-state system.\cite{leggett87, mahan90} Therefore the study of coherent tunneling regime in particle dynamics, which implies some degree of particle delocalization between wells, actually necessitates the inclusion of non-diagonal couplings which arise explicitly in these circumstances of significant wavefunction overlap.

Non-diagonal couplings have previously been studied within the context of polaron formation in molecular physics \cite{munn85b}. It has been shown that they can have a pronounced effect on the polaron properties, from a strong temperature dependent bandwidth narrowing \cite{hannewald04}, to a reduction in the phonon dressing effect \cite{stojanovic04} and an increased charge mobility \cite{munn85c}. Furthermore, non-diagonal couplings have been shown to dramatically alter a polarons' ground state properties leading to increased phonon-modulated hopping \cite{marchand17}. 

A TLS with dual-coupling to an oscillator bath has been studied within a Born-Markov approximation,\cite{wu12} as well as using a numerical approach in the ultra-strong coupling regime.\cite{acharyya20} In both cases it was found that the presence of non-diagonal system-bath couplings can lead to increased coherence times in the dynamics. Even earlier studies of a TLS with just a non-diagonal coupling found a counterintuitive relationship between the dephasing time $T_2$ and relaxation time $T_1$.\cite{laird91, reichman95} It was shown that by going beyond second order perturbation theory in the system-bath coupling, the well known equality $1/T_2 \geq 1/2T_1$ derived from the Bloch equations could actually be violated, and in fact be found to be $1/T_2 \leq 1/2T_1$ at fourth order for certain parameters and even at finite temperatures. More recently it was found, using numerical techniques, that the simultaneous presence of both non-diagonal and diagonal couplings were required in order to achieve stead-state coherences within a TLS and were remarkably independent of the initial state of the system.\cite{guarnieri18} A dual-coupling model using two independent baths---one for each interaction type---has been studied in the context of quantum phase transitions\cite{zhou15} as well as delocalised charge transfer states in organic photovoltaic settings.\cite{yao15} However, the phenomenon of enhanced coherence effects in a dual coupling model also appears be contingent on the presence of a single bath shared between both types of couplings.\cite{guarnieri18} 

In this paper we extend the SB model to include both simultaneous diagonal and non-diagonal couplings to a single oscillator bath while making use of the powerful NIBA method. This allows us to retain boson-boson cloud correlations, and facilitates a study of emergent coherence effects even at physiological temperatures and non-perturbative system bath couplings. This is distinct from techniques that invoke the Markov approximation, such as Lindblad master equation methods,\cite{breuer02} where the bath is assumed to instantaneously respond to the tunneling particle behavior and relax to its equilibrium state, thereby excluding any possible non-local (in time) correlations mediated by the environment.

It is generally assumed that coherence effects are absent in tunneling systems at relatively high temperatures and significant coupling to the bath, owing to strong thermal fluctuations in the environment. However, these conclusions are usually reached based upon models that include only diagonal couplings. Here we show that the inclusion of non-diagonal couplings can reduce decoherence rates to such a degree that the tunneling particle remains coherent for a short period of time, despite operating in a regime that would otherwise present incoherent dynamics.\cite{leggett87, weiss08} 

We study the dual-coupling model in the context of an environment characterized by an Ohmic spectral density at high temperatures $k_B T \gg \Delta$, where $\Delta$ is the tunneling energy of the TLS. Since we are interested in physiological temperatures around $T= 298K,\, k_B T=26$meV, the tunneling energies we consider are therefore of the order meV, which is relevant to molecular tunneling energy scales. Ohmic spectral densities have been used to model bosonic spin-density excitations in metals\cite{grabert92} as well as dielectric environments such as polar solvents surrounding biomolecules.\cite{gilmore08} They were also studied extensively in the original SB-model within the context of modeling superconducting flux-qubits and their environment.\cite{leggett87} 

For a TLS coupled to an Ohmic bath, NIBA is understood to be a stable approximation across the full range of coupling strengths at high-temperatures.\cite{weiss08} Conversely, NIBA is known to breakdown for the case of a biased TLS at low-temperatures, therefore our model explicitly excludes this regime where $\Delta \gg k_B T$. Nevertheless, the physiological temperatures and molecular tunneling energies considered here apply to a wide range of systems, including biological ones,\cite{adolphs06} meaning that the results presented here could help explain recent experiments\cite{engel07} on emergent macroscopic coherence effects in photosynthesis.

The rest of the paper is structured as follows. We first introduce the dual-coupling (DC) model and perform a canonical transformation to arrive at an effective DC-Hamiltonian. Then we study the dynamical properties of the central spin by averaging over the environmental degrees of freedom within a NIBA approach. We then investigate the spin dynamics demonstrating that a non-diagonal coupling introduces an additional oscillatory mode that is relatively long-lived against a background of rapid fluctuations in the environment. \footnote{Such findings are consistent with previous studies of a dual-coupling model where a separation of timescales in the coherent dynamics was found to emerge with the addition of a non-diagonal coupling mechanism\cite{acharyya20}.} We conclude that a simultaneous non-diagonal and diagonal system-bath interactions improves the coherent nature of the tunnelling particle for a short time, even in the intermediate coupling regime at relatively high temperatures.

\subsection{\label{DC_model}The dual-coupling (DC) model}
\noindent
In the dual-coupling (DC) model both the on-site energies and the tunneling energies are modulated by the bath due to the simultaneous presence of diagonal couplings to $\sigma_z$, and non-diagonal couplings to $\sigma_x$. The Pauli matrices span the Hilbert space of the central two-state tunneling system and are given by $\sigma_z=\ket{1}\!\bra{1}-\ket{2}\!\bra{2}$ and $\sigma_x=\ket{1}\!\bra{2}+\ket{2}\!\bra{1}$. The TLS can be considered a truncated form of a more general double-well potential system with extended coordinates, where the states $\ket{1},\ket{2}$ represent ground states of the isolated left and right wells respectively.\cite{leggett87} The tunneling matrix element $\Delta$, representing the tunneling frequency between the left and right wells, then mixes the two states which can also be offset in energy by the amount $\epsilon$. A general form for the Hamiltonian is given by
\begin{equation}
H \;=\; \tfrac{1}{2} [\epsilon(\{b_q\})\sigma_z  + \Delta(\{b_q\})\sigma_x]  \;+\; H_b
\label{ham_gen}
\end{equation}
where the on-site $\epsilon(\{b_q\})$ and tunneling $\Delta(\{b_q\})$ energies are functionals of the boson variables $b_q (b_q^{\dagger})$ and $H_b = \sum_q\omega_q b_q^{\dagger}b_q$ (in units of $\hbar$).

We consider an oscillator bath model where the bosonic bath is sufficiently large such that the interaction between the TLS and each bath mode is weak.\cite{feynman63, caldeira83, leggett87, weiss08, marchand17} In this case it is appropriate to consider the TLS-bath coupling to linear order and the functionals in Equation \ref{ham_gen} can be expanded to give
\begin{gather}
H = \frac{\Delta}{2} \sigma_x  + \frac{\sigma_z}{2} \sum_{q} \lambda^z_{q} (b_q+b_q^{\dagger})
+ \frac{\sigma_x}{2} \sum_q \lambda^x_{q} (b_q+b_q^{\dagger}) + H_b
\label{DC_ham}
\end{gather}
where $\lambda^{z (x)}_{q}$ is the diagonal (non-diagonal) coupling strength of each boson mode $q$ to the central spin. We also choose to consider the case of a symmetric TLS such that $\epsilon=0$.  It should be noted that here that while the linear TLS-bath coupling approximation is valid for weak coupling to each bath mode $q$, the overall interaction can be arbitrarily strong considering the fact that we have a very large number of $q$ contributing to the full coupling strength.

Going forward we introduce a parameter $\zeta=\lambda^x_q/\lambda^z_q$ that defines the strength of non-diagonal interaction relative to the diagonal interaction. We therefore consider the two forms of bath coupling to share the same bosonic frequency dependence up to a scaling factor $\zeta$.\cite{wu12} This facilitates the following analysis while retaining the essential novel physics in the model. 

It is usually assumed that a non-diagonal interaction term, of the form in Equation \ref{DC_ham}, can be neglected as it is at most a function of the wavefunction overlap between the  $\Delta$\footnote{Non-diagonal system-bath couplings change the particle wavefunction from the state $\ket{0}$ to $\ket{1}$ and vice versa. Therefore they are proportional to the overlap of the corresponding state wavefunctions $\psi_0$ and $\psi_1$ respectively in real coordinate space, indicating they are of order $\sim\Delta$ \cite{leggett87}}, and an analysis based on a NIBA approach assumes relatively small tunneling energies in the problem.\cite{leggett87} However, while we consider problems with relatively small renormalised tunneling energies, they are not negligibly small, and a retention of terms proportional to higher orders of the renormalised tunneling energy is still within the purview of the approximation. Therefore, in the following analysis we retain non-diagonal interaction terms of this kind, up to second order in the relative coupling strength parameter $\zeta$, so as to begin relaxing this assumption.\\
\indent We proceed to transform the above Hamiltonian in to the polaron frame (or shifted oscillator basis) by applying the following transformation 
\begin{equation}
\tilde{U}=e^{-S}, \quad S=(1/2)(\sigma_z+\zeta\sigma_x)\sum_q u_{q} (b_q-b_q^{\dagger})
\end{equation}
where $u_{q}=\lambda_{q}/\omega_q$. Performing the transformation $H \rightarrow \tilde{U}H\tilde{U}^{-1} \equiv \tilde{H}$, ignoring a constant energy shift as well as two boson processes,\cite{munn85b} leads to the DC-Hamiltonian
\begin{equation}
\tilde{H}=\hat{\epsilon}\sigma_z+\hat{K}_-\sigma_+ + \hat{K}_+\sigma_- + H_b
\end{equation}
where the transformed on-site energy $\hat{\epsilon}=\epsilon_{\zeta}-\delta\hat{\epsilon}$ and transformed tunneling energies $\hat{K}_{\pm}$ are now functions of the boson degrees of freedom $b,b^{\dagger}$ and are found to be
\begin{equation}
\epsilon_{\zeta}=\frac{\zeta\Delta}{2(1+\zeta^2)}, \quad\delta\hat{\epsilon} = \frac{\zeta\Delta}{2(1+\zeta^2)}\textup{cosh}(\hat{\phi}_{\zeta})
\label{onsite_energy_transformed}
\end{equation}
and 
\begin{align}
\hat{K}_{\pm} = &\frac{\Delta}{2(1+\zeta^2)}\left(\textup{cosh}(\hat{\phi}_{\zeta}) -1\right) \nonumber\\
&\pm \frac{\Delta}{2\sqrt{1+\zeta^2}}\textup{sinh}(\hat{\phi}_{\zeta})+\frac{\Delta}{2}
\end{align}
where $B_{\zeta}^{\pm} = \text{exp}(\pm \hat{\phi}_{\zeta} )$ are the bosonic shift operators \cite{mahan90, esquinazi98} and $\hat{\phi}_{\zeta} = \sqrt{1+\zeta^2}\sum_q u_{q} (b_q-b_q^{\dagger})$ and $u_q=\lambda_q/\omega_q$. 

In order to facilitate further analysis of the problem we turn to the interaction picture separating the Hamiltonian $\tilde{H}=\tilde{H}_0+\tilde{H}'$ in to $\tilde{H}_0=\epsilon_{\zeta}\sigma_z+H_b$, and the fluctuating terms given by
\begin{equation}
\tilde{H}'=\delta\hat{\epsilon}\sigma_z + \tilde{V}', \, \tilde{V}'= \hat{K}_-(t)e^{i\epsilon_{\zeta} t}\tilde{\sigma}_+ + \hat{K}_+(t)e^{-i\epsilon_{\zeta} t}\tilde{\sigma}_-
\label{int_pic_ham}
\end{equation}
where we are now in a frame rotating at the renormalised transition frequency $\epsilon_{\zeta}$. 

The time evolution of the boson operators $b_q(t)=e^{-i\tilde{H}_0 t}b_q e^{i\tilde{H}_0 t}$ is independent of the coupling to the central spin and are given by $b_q(t)=e^{-i\omega_q t}b_q$, $b_q^{\dagger}(t)=e^{i\omega_q t}b_q^{\dagger}$. The bath displacement operators therefore evolve in time according to
\begin{equation}
B_{\zeta}^{\pm}(t)=\text{exp}\left[\pm \sqrt{1+\zeta^2}\sum_q u_{q} (b_q e^{-i\omega_q t}-b_q^{\dagger}e^{i\omega_q t})\right]
\label{disp_op}
\end{equation}
Before continuing we briefly remark on the limiting cases of pure diagonal and non-diagonal couplings separately, to help us better understand their individual effects on the spin dynamics.

For pure diagonal coupling, the Hamiltonian reduces to $H_{SB}=(\Delta/2)\sigma_x+(\sigma_z/2)\sum_q\lambda^z_q(b_q+b_q^{\dagger})+H_b$, in which case we recover the unbiased SB model and all of its results. Here the particle can tunnel between degenerate states of the TLS subject to fluctuating energy levels arising due to boson emission/absorption. The bosonic bath is effectively `measuring' the state of the TLS leading to a loss of phase coherence as well as energy loss to the environment.\cite{weiss08} In the high-temperature limit $k_b T \gg \Delta$, the TLS is understood to undergo incoherent tunneling for most bath spectral density functions that have been investigated analytically, and for all but the smallest system-bath coupling strengths.

For pure non-diagonal coupling, the Hamiltonian reduces to $H_{IB}=(\Delta/2)\sigma_z+(\sigma_z/2)\sum_q\lambda^x_q(b_q+b_q^{\dagger})+H_b$, where we have performed a rotation of the Pauli operators $\sigma_x \rightarrow \sigma_z$ to recover the well known Independent-Boson Hamiltonian (IB).\cite{mahan90} We now have a biased (or asymmetric) TLS subject to a fluctuating polarization field. $H_{IB}$ is an exactly solvable model with a simple canonical transformation leading to $\tilde{H}_{IB}=\sigma_z(\Delta-\Omega)/2+H_b$, where $\Omega=\sum_q(\lambda_q^x)^2/\omega_q$. In this case we see how the non-diagonal coupling affects the tunneling process leading to a renormalization of the tunneling energy by an amount $\Omega$. 

Finally we would like to remark on the possibility of applying a rotation to $H$, specifically in the $\sigma_y$ direction, to recover a diagonal coupling with the appropriate choice of new axes. Such a trick would of course remove the troublesome non-diagonal coupling albeit at the expense of introducing a bias term to the central spin system as well as rotating the basis in to a linear superposition. While a biased tunneling system with a diagonal bath coupling has been previously studied, the NIBA approach was formulated in the $\sigma_z$ basis of the system. As we discuss further below, NIBA is valid for short `excursions' of the system dynamics in to the off-diagonal elements of the density matrix. Provided one is working in the $\sigma_z$ basis, such off-diagonal terms represent the coherences in the system and NIBA remains a good approximation for systems subject to strong environmental fluctuations where coherences are relatively short lived. In a rotated basis the diagonal and off-diagonal density matrix no longer carry the same meaning and NIBA is no longer justified, as off-diagonal `excursions' become relatively long-lived and correlations between these events become long range in time. 

We therefore compare our results for the DC model not only to the biased SB model within NIBA, but also to a well known extension to the model known as the nearest-neighbor-blip approximation (NN).\cite{weiss08} In this case  longer lived coherences in the model are permitted, leading to robust oscillations in the system dynamics over time that would be otherwise absent in the SB model with just NIBA. NN therefore gives us a another benchmark model with which to compare the distinct effects introduced by the DC model.

\subsection{Dynamics in the DC model}
\noindent
We wish to study the time evolution of the TLS under the influence of the oscillator bath. This means that we are interested in tracking the degrees of freedom of the TLS while averaging over (or `integrating out') the boson degrees of freedom.\cite{feynman63,feynman10} We choose to analyse the quantity $\langle\sigma_z(t)\rangle_b$, which is often most easily measured in experiment\footnote{One could of course choose to study the quantity $\sigma_x(t)$ instead which would facilitate a calculation of the dephasing time $T_2$ for example. In this case, one would have to be careful to apply the appropriate rotation to the initial state and calculate all the resulting correlation functions that remain within NIBA and the small $\zeta$ limit}. This averaging process amounts to a partial trace over the bath degrees of freedom, which is represented by $\langle...\rangle_b=\text{Tr}_b(\rho_b..)$, where the initial global density matrix has been assumed to be factorisable such that $\rho(0)=\rho_s\rho_b$\footnote{The condition $\rho(0)=\rho_s\rho_b$ can be physically realised in situations where the spin system is held in a certain position, for e.g by a strong bias field, such that the bath can come in to thermal equilibrium with it. Then at $t=0$, the system is released, and the subsequent dynamics are governed by the Hamiltonian $H$\cite{leggett87}. This approach can can also be used to model physical systems under similar initial conditions whereby the spin state is (effectively) instantaneously prepared, such as photon absorption, or electron injection, in a molecular setting}, and $\rho_b=e^{-\beta H_b}/\text{Tr}(e^{-\beta H_b})$.\cite{breuer02} 

Since we wish to operate in the polaron frame, we insert the identities $\mathds{1}=\tilde{U}^{\dagger}\tilde{U}$. This produces the desired transformed operator $\tilde{\sigma}_z(t)=\tilde{U}^{\dagger}\sigma_z(t)\tilde{U}$, whose dynamics are governed by $\tilde{H}$, but also acts on the initial spin state which we assume is $\rho_s=\ket{1}\!\bra{1}$. The time-domain spin polarization is thus
\begin{equation}
\langle \sigma_z (t)\rangle_b=\big\langle \tilde{\sigma}_z(t)\big(\tilde{U}^{\dagger}\rho_s\tilde{U}\big) \big\rangle_b
\end{equation}
where we have applied the cyclic properties of the trace. Transforming the initial state of the spin produces
\begin{align}
\langle \sigma_z (t)\rangle_b=&\langle \tilde{\sigma}_z(t)\rho_s\rangle_b+\langle \tilde{\sigma}_z(t)\tilde{\sigma}_z\rangle_b+\langle \tilde{\sigma}_z(t)\tilde{\sigma}_+\rangle_b +\langle \tilde{\sigma}_z(t)\tilde{\sigma}_-\rangle_b
\label{sigmaz_rotated}
\end{align}
where
\begin{align}
&\tilde{\sigma}_z=\sigma_z\frac{\zeta^2}{2(1+\zeta^2)}(1-\cosh(\hat{\phi}_{\zeta}))\nonumber\\
&\tilde{\sigma}_+=\sigma_{+}\left[\frac{\zeta}{2(1+\zeta^2)}\left(\cosh(\hat{\phi})_{\zeta}-1\right)-\frac{\zeta}{2\sqrt{1+\zeta^2}}\sinh(\hat{\phi})_{\zeta}\right]\nonumber\\ &\tilde{\sigma}_-=\sigma_{-}\left[\frac{\zeta}{2(1+\zeta^2)}\left(\cosh(\hat{\phi})_{\zeta}-1\right)+\frac{\zeta}{2\sqrt{1+\zeta^2}}\sinh(\hat{\phi}_{\zeta})\right]
\label{sigmaz_rotated}
\end{align}
In order to proceed we turn to the interaction picture. The spin operators currently evolve in time in the Heisenberg picture $\tilde{\sigma}_z(t)=e^{i\tilde{H}t}\tilde{\sigma}_z e^{-i\tilde{H}t}$. Recasting Equation \ref{sigmaz_rotated} with operators defined in the interaction picture $\tilde{\sigma}^{I}_z(t)=e^{i\tilde{H}_0t}\tilde{\sigma}_z e^{-i\tilde{H}_0t}$
\begin{align}
&\langle\sigma_z(t)\rangle_b=\big\langle\tilde{U}'^{\dagger}(t)\tilde{\sigma}_z^I(t)\tilde{U}'(t)\rho_s\big\rangle_b+\big\langle\tilde{U}'^{\dagger}(t)\tilde{\sigma}_z^I(t)\tilde{U}'(t)\tilde{\sigma}_z\big\rangle_b\nonumber\\
&+\big\langle\tilde{U}'^{\dagger}(t)\tilde{\sigma}_z^I(t)\tilde{U}'(t)\tilde{\sigma}_+\big\rangle_b + \big\langle\tilde{U}'^{\dagger}(t)\tilde{\sigma}_z^I(t)\tilde{U}'(t)\tilde{\sigma}_-\big\rangle_b
\label{sigmaz_rotated_interaction}
\end{align}
the time-evolution operators $\tilde{U}'(t)=e^{i\tilde{H}_0t}e^{-i\tilde{H}t}$ and the interaction picture Hamiltonian $\tilde{H}'$ defined in Equation \ref{int_pic_ham}. 

The Greens functions are defined in the interaction picture via $\tilde{G}_{\alpha\beta}=\bra{\alpha}\tilde{U}'(t)\ket{\beta}$ and represent the probability amplitudes, tracking all the possible paths the system might take. In their infinite series representation they are given by
\begin{align}
&\tilde{G}_{\alpha\beta}(t)=\sum_{n=0}^{\infty}(-i)^{n}\int_0^{t}d\tau_{n}...\!\!\int_0^{\tau_2}d\tau_1 \nonumber\\
&\times\bra{\alpha}\Big[\delta\hat{\epsilon}(\tau_{n})\sigma_z+e^{-i\epsilon_{\zeta}t}\hat{K}_+(\tau_{n})\sigma_-+e^{i\epsilon_{\zeta}t}\hat{K}_-(\tau_{n})\sigma_+\Big]\nonumber\\
&\times...\Big[\delta\hat{\epsilon}(\tau_{1})\sigma_z+e^{-i\epsilon_{\zeta}t}\hat{K}_+(\tau_{1})\sigma_-+e^{i\epsilon_{\zeta}t}\hat{K}_-(\tau_{1})\sigma_+\Big]\ket{\beta}
\label{transition_greens_inf_series_non_local}
\end{align}
In the polaron frame Equation (\ref{transition_greens_inf_series_non_local}) intuitively describes a particle tunneling back and forth between the ground states of each potential well, while simultaneously dragging a cloud of bosons with it. The non-diagonal bath coupling has introduced the additional complexity of $\hat{\epsilon}$ acting in between flips while the particle dwells in each state, as well as complicated functions $\hat{K}_{\pm}$ of the bath displacement operators $\hat{B}_{\pm}$ acting during tunneling events. 

Within a NIBA approach one considers only short-lived coherences in the system which, in its original path integral formalism, amounts to a suppression of paths visiting the off-diagonal portions of the density matrix.\cite{leggett87} Using a polaron transformation method (as we have done here), it has been shown that such an approximation amounts to considering only nearest-neighbor (in time) bath fluctuations.\cite{esquinazi98, dekker87, weiss08, wurger97b, wurger98} Physically, this means a boson cloud $\hat{B}_{\pm}$ created at time $t_n$ only interacts with a previous cloud created at $t_{n-1}$, and all other correlations beyond this `die out' in the relevant time-frame. In its original formulation,\cite{leggett87} NIBA was argued to be valid in the limit of small $\Delta/\omega_c$, which is the ratio of tunneling energy to the characteristic frequency of the bath modes. Furthermore, the non-diagonal interaction strength is of the order $\mathcal{O}(\Delta)$ given that it corresponds to the wavefunction overlap between the states comprising the TLS.\cite{leggett87} It is therefore consistent within a NIBA approach to consider only small non-diagonal couplings up to $\zeta^2$.

%\textcolor{green}{We will also be using NN which goes beyond NIBA by taking in to account correlations between neighbouring 'blips', which in the language presented here involves combinations of consecutive boson clouds. Therefore NN allows for slightly longer lived correlations between boson clouds.}

With these considerations in mind we retain only nearest-neighbour phonon cloud correlations due to tunneling events---which take the form $\langle \hat{K}_{+}(t_n)\hat{K}_{-}(t_{n-1})\rangle_b$. Correlations between tunneling and on-site events are of order $\langle \hat{K}_{+}(t_n)\delta\hat{\epsilon}(t_{n-1})\rangle_b \sim \mathcal{O}(\zeta\Delta^2/\omega_c^2)$, while those between successive on-site events are of order $\langle \delta\hat{\epsilon}(t_n)\delta\hat{\epsilon}(t_{n-1})\rangle_b \sim \mathcal{O}(\zeta^2\Delta^2/\omega_c^2)$, and are therefore suppressed. 

Our final estimation of the spin polarization (see Appendix \ref{spin_pol_est_appendix}) therefore reduces to
\begin{align}
\langle\sigma_z(t)\rangle_b\approx\langle \tilde{G}_{11}^{\dagger}(t)\tilde{G}_{11}(t)\rangle_b
\end{align}
Expanding the Greens functions according to their infinite series representation and retaining nearest-neighbour bath fluctuations allows us to re-sum the series as
\begin{align}
&\langle\sigma_z(t)\rangle_b= 1-\tfrac{1}{2}\int_0^t \!\!d\tau \int_0^{\tau} \!\!d\tau' \,\,\Sigma_{\text{DC}}(\tau,\tau')\sigma_z (\tau')
\label{rho_expanded} 
\end{align}
with the kernel, or self-energy, acting between times $\tau$ and $\tau'$ given by
\begin{gather}
\Sigma_{\text{DC}}(\tau-\tau') = e^{i\epsilon_{\zeta}(\tau-\tau')}\Big\langle K_+(\tau)K_-(\tau') \Big\rangle_b \nonumber\\
+ e^{-i\epsilon_{\zeta}(\tau-\tau')}\Big\langle K_-(\tau)K_+(\tau') \Big\rangle_b
\label{kernel_approx}
\end{gather}
The bath correlation functions contained in the expansion of Equation \ref{kernel_approx} can be evaluated using a Feynman operator disentangling method,\cite{mahan90} and expressed in the continuum limit as
\begin{gather}
\langle B_{\pm}(t) B_{\mp}(0) \rangle_b = \langle B_{\pm} \rangle \langle B_{\mp} \rangle e^{\varphi(t)}
\end{gather}
where the phase
\begin{align}
\varphi(t)&= iQ'(t)-Q''(t)
\end{align}
is comprised of the well known bath correlation functions\cite{weiss08}
\begin{align}
&Q'(t)=\int_0^{\infty}\!\!d\omega \frac{J(\omega)}{\omega^2}\sin(\omega t) \nonumber\\
&Q''(t)=\int_0^{\infty}\!\!d\omega \frac{J(\omega)}{\omega^2}(1-\cos(\omega t))\coth(\beta\omega/2)
\label{bath_correlation_functions}
\end{align}
The spectral density function $J(\omega)=(\pi/2)\sum_q (\lambda^2_{q}/\omega_q)\delta(\omega-\omega_q)$ describes the distribution of bath modes weighted by their coupling to the central spin. 

The self-energy $\Sigma_{\text{DC}}(\tau,\tau')=\Sigma_{\text{DC}}(\tau-\tau')$ is then a function of the time difference $\tau-\tau'$ for the oscillator bath in question. The system can therefore be solved by taking its Laplace transform $\rho(\lambda)=\int_0^{\infty}dt\, e^{-\lambda t}\rho(t)$, to produce
\begin{equation}
\langle\sigma_z(\lambda)\rangle_b= \frac{1}{\lambda+\Sigma_{\text{DC}}(\lambda)}
\label{den_mat_lambda}
\end{equation}
\noindent
We are now left with evaluating the Laplace transform of the kernel function
\begin{align}
\Sigma_{\text{DC}}(\lambda)&=\frac{\Delta^2}{2}\int_0^{\infty}\!\!dt e^{-\lambda t}\cos(\epsilon_{\zeta} t)\Bigg[I -\frac{\sinh\left(\varphi(t)\right)}{1+\zeta^2}\nonumber\\
&+\frac{\cosh\left(\varphi(t)\right)}{(1+\zeta^2)^2} \Bigg]
\label{DCSB_self_energy}
\end{align}
where
\begin{equation}
I=\frac{1-2\mathbb{B}}{(1+\zeta^2)^2}-\frac{2(1-\mathbb{B})}{1+\zeta^2}+1
\end{equation}
The terms $\mathbb{B} \equiv \langle B_{\pm} \rangle_b$ are the temperature-dependent Franck-Condon factors, which can be cast in the continuum limit as\cite{weiss08}
\begin{equation}
\mathbb{B} = \text{exp}\Big[-\int_0^{\infty} d\omega \frac{J(\omega)}{\omega^2}\coth(\beta \omega/2)\Big].
\end{equation}
and represent the renormalization of the tunneling terms due to boson cloud `dressing effect'.

For the corresponding NN expressions for the spin polarization and self-energies, see Appendix \ref{NNBA_appendix}.

\subsection{Results for the Ohmic-DC model}
\noindent
It now remains to define the form of the bath spectral density. Here, we use an Ohmic form given by
\begin{equation}
J(\omega)= \eta\omega e^{-\omega/\omega_c}
 \label{Jw_DCSB_ohmic}
\end{equation}
where $\omega_c$ is a characteristic frequency of the bath and $\eta$ is the friction coefficient.\cite{weiss08} The dimensionless measure of the diagonal coupling parameter is therefore $\gamma=\eta /\hbar$ (where we have reintroduced $\hbar$ for clarity) \footnote{For a bath consisting of conduction electrons in a solid at low-temperatures, it has been shown that the corresponding charge-density excitations follow an Ohmic form for the spectral density \cite{leggett87}. Therefore, in the relevant limit, the parameter $\gamma$ for the dimensionless TLS-bath coupling is equivalent to the Kondo parameter \cite{kondo84}, indicating that the model can be used to describe a magnetic spin-1/2 impurity interacting with an electron gas.}. 

Using standard results\cite{leggett87, weiss08} for the Ohmic bath, the bath correlation functions in Equation \ref{bath_correlation_functions} are now given by
\begin{align}
&Q'(t)=\gamma\tan^{-1}\omega_c t\nonumber\\
&Q''(t)=\frac{\gamma}{2}\text{ln}(1+\omega_c^2 t^2)+\gamma\text{ln}\left[\frac{\beta\hbar}{\pi t}\text{sinh}\left(\frac{\pi t}{\beta\hbar}\right)  \right]
\label{correlation_function_ohmic_t}
\end{align}
One can gain some appreciation at this stage as to the validity of NIBA for an Ohmic bath in the high-temperature limit.  Correlations between long-range bath fluctuations contain functionals $\text{exp}[-Q''(\delta t)]$ with relatively large time intervals compared to $\beta\hbar$, producing vanishingly small contributions to the kernel $\Sigma_{DC}$ and can therefore be discarded within NIBA\footnote{For in an depth discussion around the validity of NIBA for Ohmic baths at high-temperatures see \cite{leggett87} and \cite{weiss08}}. 

%
%The Ohmic bath correlation functions can now be seen to be increasing for long times and high-temperatures. Correlations between bath fluctuations separated by long time intervals are 
%
%One can see more clearly now why NIBA is justified in the high-temperature limit for an Ohmic bath. For small $\beta$, correlations between bath fluctuations separated by long times contain functions $Q''(t)$ that increase linearly with $t$. Long range bath correlations (or blips) 

Evaluating the high-temperature Franck-Condon factors based on a variational approach \cite{silbey85a} gives
\begin{equation}
\mathbb{B} = (\pi\mu\omega_c)^{-\gamma}
\end{equation}
and we find the kernel to be\footnote{We have made use of the fact that for an Ohmic bath (see Eq. \ref{correlation_function_ohmic_t}), $Q''(t,\beta)=Q''(t,-\beta)$ and therefore $-Q''(t)=Q''(t)$}
\begin{widetext}
\begin{align}
\Sigma_{\text{DC}}(\lambda) = &\frac{\mathbb{B}^2\Delta^2 \mu \Gamma(1-2\gamma\sqrt{1+\zeta^2})}{2} \Bigg\{\frac{\cos(\pi \gamma\sqrt{1+\zeta^2})}{(1+\zeta^2)^2}\Bigg[f(\lambda+i\epsilon_{\zeta})+f(\lambda-i\epsilon_{\zeta})\Bigg] \nonumber\\
&-i\frac{\sin(\pi \gamma\sqrt{1+\zeta^2})}{(1+\zeta^2)}\Bigg[f(\lambda+i\epsilon_{\zeta})-f(\lambda-i\epsilon_{\zeta})\Bigg]\Bigg\}+\frac{I \Delta^2 \lambda}{\lambda^2+\epsilon_{\zeta}^2}
\end{align}
\end{widetext}
where
\begin{equation}
f(\lambda)=\frac{1}{\gamma\sqrt{1+\zeta^2}+\mu\lambda}\frac{\Gamma(\gamma\sqrt{1+\zeta^2}+\mu\lambda)}{\Gamma(1-\gamma\sqrt{1+\zeta^2}+\mu\lambda)}
\label{inf_fun_high_T_ohm}
\end{equation}
and $\Gamma(z)$ is the Gamma function.

We are now in a position to present results for the dynamics of the DC model. The time-domain spin polarization is recovered by taking its inverse Laplace transform: $ \langle \sigma_z(t) \rangle=(1/2\pi i)\int_C d\lambda e^{\lambda t}\langle \sigma_z(\lambda) \rangle$. As it stands, Equation \ref{inf_fun_high_T_ohm} produces a pole structure to $\langle \sigma_z(t) \rangle$ to all orders in $\lambda$, and therefore must be approximated in order to facilitate contour integration required by the inverse Laplace transform. In the interest of studying the crossover from coherent to incoherent motion, we consider the high-temperature limit $\mu=1/(2\pi k_b T) \ll \Delta$ allowing us to expand the self-energy to second order in $\mu\lambda$. Quoting standard results,\cite{leggett87} the function $f(\lambda)$ can be approximated in the high-T limit as
\begin{align}
f(\lambda) = \frac{\mu\,\nu}{\gamma+\mu\lambda}\left( 1+\mu\Lambda\lambda+\frac{\mu^2\Theta}{2}\lambda^2 \right)
\end{align}
where
\begin{align}
\nu&=\cos(\pi\gamma)\frac{\Gamma(1+\gamma)\Gamma(1-2\gamma)}{\Gamma(1-\gamma)}, \nonumber\\
\Lambda&=\psi_0(1+\gamma)-\psi_0(1-\gamma) \nonumber\\
\Theta&=\psi_0(1-\gamma)^2-2\psi_0(1-\gamma)\psi_0(1+\gamma)\nonumber\\&-\psi_0(1+\gamma)^2+\psi_1(1+\gamma)
\end{align} 
The functions $\Gamma(z), \psi_0(z), \psi_1(z)$ are the gamma function, digamma function and trigamma function respectively.
\begin{figure}[h]
\includegraphics[width=0.45\textwidth]{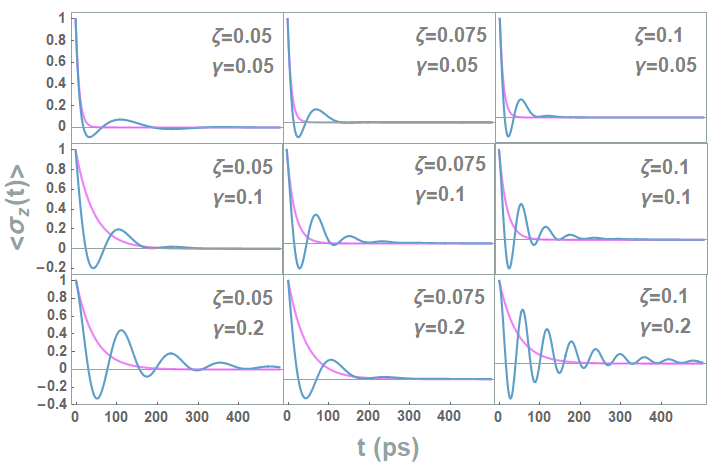}
				\caption{Spin polarization dynamics for DC model (blue) compared to SB model within NIBA (pink, $\zeta=0$). Parameters: $k_b T=26$meV, $\Delta=1$meV, $\hbar \omega_c=100$meV.}
		\label{P_11_t}
\end{figure}
\\
\indent In Figure \ref{P_11_t} we plot the time-domain spin dynamics. For zero non-diagonal coupling strength $\zeta=0$ we recover the SB model with only diagonal coupling (shown in pink for each plot). For all but the smallest coupling strengths $\gamma$, the SB model exhibits overdamped motion with no coherent oscillations. This is consistent with the behavior of the Ohmic-SB model at high temperatures relative to the tunneling energy which is expected to exhibit incoherent tunneling.\cite{leggett87, weiss87, shnirman02} This is owing to strong thermal fluctuations in the bath at the physiological temperatures $T=298K$ considered here. In Figure \ref{P_11_t}, for $\zeta=0$, we also see the manifestation of the quantum Zeno effect;\cite{harris82} the process by which the environment rapidly `measures' the state of the central spin, tending towards localising it in its initial state. This is characterised in $\langle \sigma_z(t)\rangle$ by an increase in relaxation time as a function of the coupling strength.\cite{shnirman02}

For the DC model, as $\zeta$ is switched on we begin to observe a revival in coherent oscillations. For small coupling strength $\gamma=0.05$, this effect is small and the oscillations die off rapidly over the course of $\sim$ 100ps. For larger $\gamma = 0.1$, we move out of the perturbative regime and in to the intermediate coupling regime for $\gamma \geq 0.1$. In this regime we begin to observe more pronounced coherent oscillations that persist for many hundreds of picoseconds even for relatively weak coupling to the bath. 

\begin{figure}[h]
\subfloat[]{\includegraphics[width=0.23\textwidth]{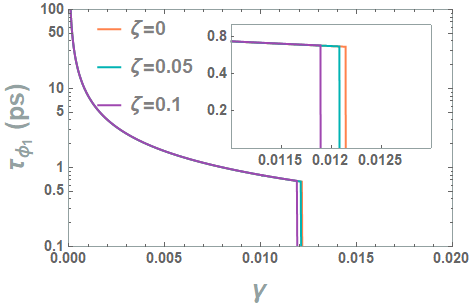}}
\subfloat[]{\includegraphics[width=0.23\textwidth]{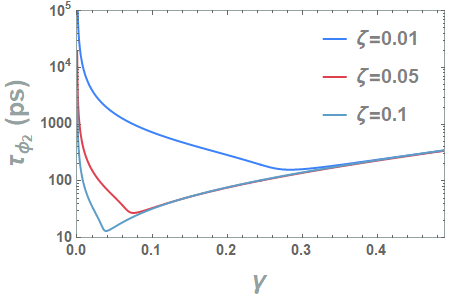}}
				\caption{Coherence times of tunneling particle vs dimensionless bath coupling strength $\gamma$ for various $\zeta$. Parameters: $k_b T=26$meV, $\Delta=1$meV, $\hbar \omega_c=100$meV.}
		\label{dynamical_props}
\end{figure}

The improvements in coherence times observed in Fig. \ref{P_11_t} can be quantified by considering the dynamical properties of the tunneling particle, which are determined by the pole structure of $\langle \sigma_z(t) \rangle$ in Equation \ref{den_mat_lambda}. Upon Laplace inversion of the spin polarization, entirely real poles lead to pure relaxation driving the system towards its equilibrium value, while the complex conjugate poles determine the duration and frequency of coherent oscillations. We therefore interpret the coherence times as $\tau_{\phi_{i}}=1/\textup{Re}(\lambda_{i})$, which defines the timescale over which the oscillatory components are suppressed for a given mode $i$ oscillating with frequency $\textup{Im}(\lambda_{i})$.
In the SB model there exists only one oscillatory mode $\tau_{\phi_{1}}$. In Fig \ref{dynamical_props} (a) the coherence time of this mode is indicated by the orange line, and the expected coherent-to-incoherent transition\cite{leggett87,weiss87} is observed for relatively small $\gamma\sim 0.012$. For the DC model, the coherence time $\tau_{\phi_{2}}$ of the additional oscillatory mode is plotted in Figure \ref{dynamical_props} (b), for various non-diagonal coupling strengths $\zeta$. In addition, the effect of $\zeta$ on $\tau_{\phi_{1}}$, is plotted in \ref{dynamical_props} (a) for $\zeta=0.05$ (green line) and $\zeta=0.1$ (purple line).
The effect of $\zeta$ on $\tau_{\phi_{1}}$ is relatively small, with the predominant effect being a slight shift of the coherent-incoherent transition line towards smaller coupling strengths, while the longevity of this coherent mode remains largely unchanged below the transition point. Conversely, the coherence time $\tau_{\phi_{2}}$ of the new coherent mode introduced by the DC model displays a markedly different $\gamma$ dependence. Not only is a coherent-incoherent transition absent across the entire region of the coupling strengths explored here $0 < \gamma < 0.5$, but the inverse relationship between $\tau_{\phi}$ and $\gamma$ is absent for intermediate coupling strengths. Instead we see a steady increase in coherence times of this mode with $\gamma$ in the non-perturbative coupling regime.

\begin{figure}[h]
\subfloat[]{\includegraphics[width=0.5\textwidth]{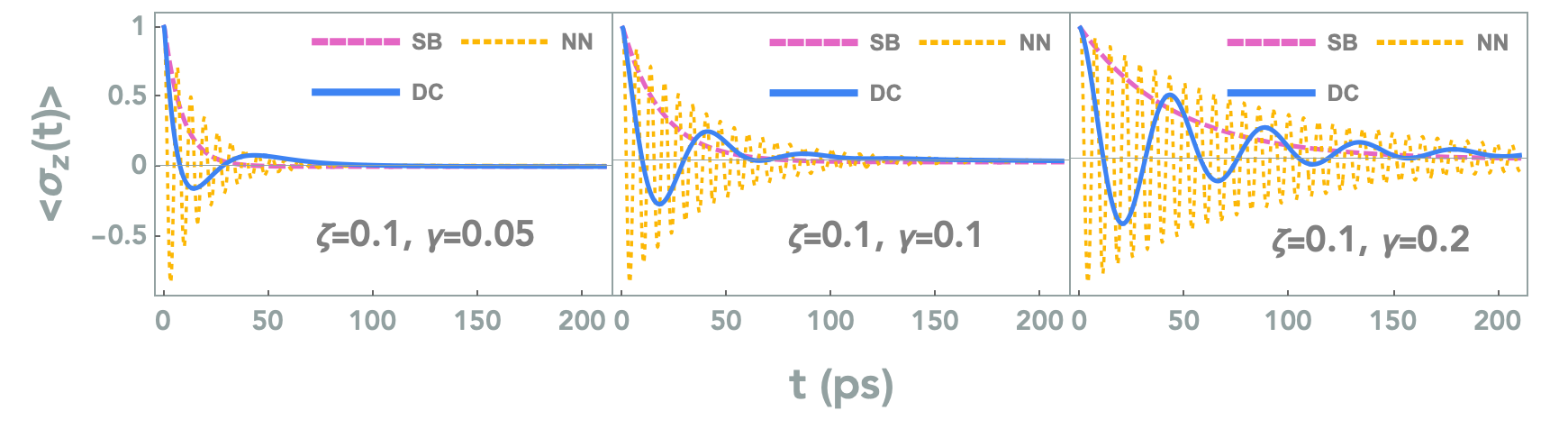}}\\
\subfloat[]{\includegraphics[width=0.3\textwidth]{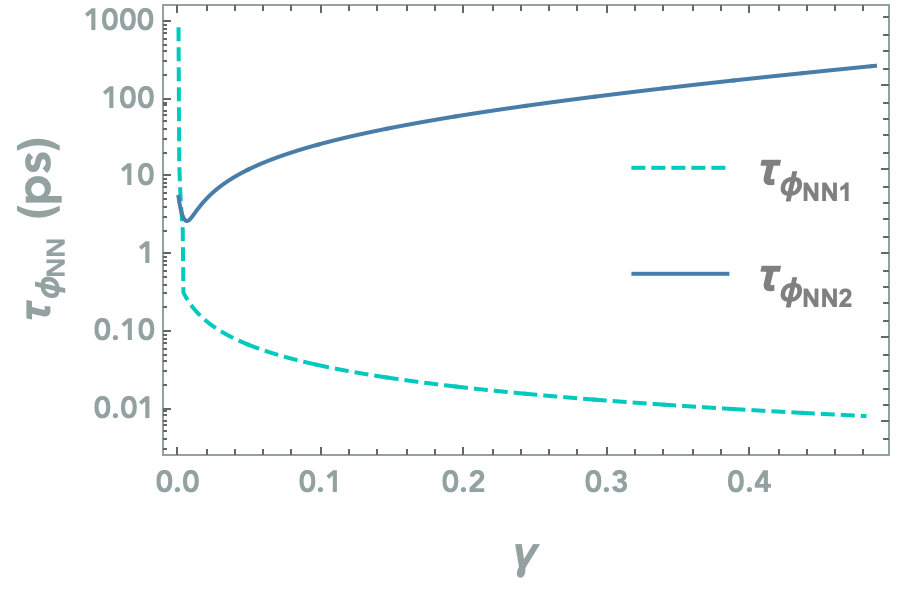}}
				\caption{Spin polarization dynamics for DC model compared to NN model (a). Coherence times $\tau_{\phi_{\text{NN1}}}$, $\tau_{\phi_{\text{NN2}}}$ of NN model (b). Parameters: $k_b T=26$meV, $\Delta=1$meV, $\hbar \omega_c=100$meV.}
		\label{P_11_t_NN}
\end{figure}

In Fig. \ref{P_11_t_NN}(a) we compare the spin polarization dynamics of the NN model to those of the DC and SB model. In the NN model, `nearest-neighbor blips' are retained which---in the formulation of the original SB model---corresponds to longer lived tunneling particle coherence. This is evident in Fig. \ref{P_11_t_NN}(a) with the longer lived oscillations in the NN model compared to its SB counterpart with only NIBA. Furthermore in Fig. \ref{P_11_t_NN}(b), we see two oscillation frequencies, with their associated relaxation times $\tau_{\phi_{\text{NN1}}}$, $\tau_{\phi_{\text{NN2}}}$, emerge for the NN model also. Compared to the DC model, we see that the NN model exhibits much faster oscillations but shorter coherence times for the long-lived modes (see Fig. \ref{P_11_t_NN}(b) compared to Fig. \ref{dynamical_props}(b)). 

To help understand these results, consider the physical mechanism introduced by the fluctuating $\sigma_x$ term in Equation \ref{DC_ham} representing the non-diagonal coupling. This term in the Hamiltonian permits boson absorption and emission processes, allowing the particle to tunnel between states via particle exchange mediated by the bath. This introduces an additional tunneling mechanism for the particle beyond just the $\Delta\sigma_x$ term in the central spin Hamiltonian. Therefore, with a non-diagonal bath coupling we might expect some degree of tunneling enhancement, within certain regions of the parameter space, owing to the additional `tunneling pathways' introduced by boson emission/absorption. 

In position space this form of coupling arises due to the bath modulating the separation distance between tunneling centers, for e.g. defect tunneling in solids whereby the host atoms or molecules are moved closer together or farther apart due to vibrational modes in the surrounding medium. In this picture it is intuitive to see how under certain circumstances, vibrations in the environment can cause a reduction in distance between tunneling centers, thereby increasing the probability of the particle tunneling.

For a TLS coupled to a very large environment, with many degrees of freedom, one expects any energy associated with an initial excitation of the spin to eventually be lost to the bath. Furthermore, the rapid fluctuations present in a thermal bath at physiological temperatures are expected to quench quantum coherence effects in the central spin system owing to high-frequency measurements of the quantum state of the system. These effects are characterised in the central spin dynamics by a rapid relaxation of the spin to its equilibrium value as well as the suppression of any oscillatory components to the dynamics. 

While our results for the DC model are qualitatively consistent with these expectations, the addition of a non-diagonal coupling mechanism is shown to extend the duration and amplitude of coherent oscillations in the TLS, even in the presence of significant environmental `noise' at physiological temperatures. What we have shown in this work is how these effects can be mitigated, to some extent, by the inclusion of a non-diagonal coupling mechanism that enhances the particle tunneling probability enough to maintain coherent oscillations, albeit for a short duration, in a system that would otherwise be in the entirely incoherent regime of tunneling. The DC model was also shown to introduce coherent oscillations comparable to the NN extension to NIBA, albeit with lower oscillation frequencies in the spin polarization. This suggests that oscillatory modes within the DC model are at least as robust as those introduced by the NN model which assumes longer lived coherences from the outset.

\subsection{\label{Summary}Summary}
\noindent We have studied an unbiased, two-state (spin 1/2) system in the presence of simultaneous diagonal and non-diagonal couplings to a shared Ohmic oscillator bath at molecular tunneling energy scales and physiological temperatures ($T=298K$). 

We have shown how the presence of non-diagonal system-environment interactions alter the coherent dynamics of the tunneling particle, leading to a revival in coherent oscillations despite strong thermal fluctuations from the bath. Contrary to the assumption that tunneling systems, operating at relatively high temperatures and significant bath couplings should present incoherent dynamics, we demonstrate that the commonly neglected non-diagonal coupling may actually help preserve coherence effects for short durations. This is evidenced by prolonged and amplified coherent oscillations observed in the tunneling particle dynamics associated with an additional, relatively long-lived oscillatory mode, introduced by the non-diagonal coupling mechanism. 

The approach used in this work, which is based on a polaron transformation method and the noninteracting-blip approximation, sets the stage for further investigation in to the deep coherent regime with extensions to the model such as the nearest-neighbor-blip approximation. Such an approximation would permit boson cloud correlations that are further non-local in time and thereby retain coherence effects in the model to higher orders. This would allow investigation of intermediate temperatures and couplings not explored in this work. 

Furthermore, investigation in to the zero and low temperature regimes could reveal a potential enhancement to the long-lived coherence effects observed in the model, which would be consistent with recent predictions regarding the low-temperature enhancement of `steady-state coherences' in a dual-coupling model.\cite{guarnieri18} Initial state preparation, by way of a shift to the equilibrium position of the oscillators, has also been shown to influence oscillatory dynamics in a strong coupling setting\cite{acharyya20} and could therefore enhance these effects even further.

We would like to thank M. Berciu and A. Nocera for many helpful discussions and critical readings of the manuscript. This work was funded as part of the Grand Challenges project at the Stewart Blusson Quantum Matter Institute at the University of British Columbia.

\normalem
\bibliography{./references}
*Electronic address: leon.ruocco@ubc.ca
\bibliographystyle{unsrt}

\appendix
\section{Approximation of the spin polarization in the rotated polaron frame}
\label{spin_pol_est_appendix}

Starting from the interaction picture spin polarization in the rotated polaron frame in Equation \ref{sigmaz_rotated_interaction} and performing the trace over spin degrees of freedom gives
\begin{align}
&\langle\sigma_z(t)\rangle_b=\big\langle \bra{1}\tilde{U}'^{\dagger}(t)\tilde{\sigma}^I_z(t)\tilde{U}'(t)\ket{1}\big\rangle_b\nonumber\\
&+\big\langle\bra{1}\tilde{U}'^{\dagger}(t)\tilde{\sigma}^I_z(t)\tilde{U}'(t)\ket{1}(\zeta\hat{\epsilon}/\Delta)\big\rangle_b\nonumber\\
&-\big\langle\bra{2}\tilde{U}'^{\dagger}(t)\tilde{\sigma}^I_z(t)\tilde{U}'(t)\ket{2}(\zeta\hat{\epsilon}/\Delta)\big\rangle_b\nonumber\\
&+\big\langle\bra{1}\tilde{U}'^{\dagger}(t)\tilde{\sigma}^I_z(t)\tilde{U}'(t)\ket{2}(\zeta/\Delta)(\hat{K}_- -\tfrac{\Delta}{2})\big\rangle_b \nonumber\\
&+\big\langle\bra{2}\tilde{U}'^{\dagger}(t)\tilde{\sigma}^I_z(t)\tilde{U}'(t)\ket{1}(\zeta/\Delta)(\hat{K}_+ -\tfrac{\Delta}{2})\big\rangle_b
\end{align}
where $\tilde{\sigma}^I_z(t)\rightarrow \sigma_z$ since $\sigma_z$ commutes with $\tilde{H}_0$. 

The Greens functions are defined in the interaction picture via $\tilde{G}_{\alpha\beta}=\bra{\alpha}\tilde{U}'(t)\ket{\beta}$ and given the initial condition $\langle \sigma_z(0) \rangle =1$, the spin polarization simplifies to
\begin{align}
&\big\langle\sigma_z(t)\big\rangle_b=\big\langle \tilde{G}_{11}^{\dagger}(t)\tilde{G}_{11}(t)\big\rangle_b \nonumber\\
&+\big\langle \tilde{G}_{11}^{\dagger}(t)\tilde{G}_{11}(t)(\zeta\hat{\epsilon}/\Delta)\big\rangle_b\nonumber\\
& - \big\langle \tilde{G}_{21}^{\dagger}(t)\tilde{G}_{12}(t)(\zeta\hat{\epsilon}/\Delta)\big\rangle_b\nonumber\\
&+\big\langle \tilde{G}_{11}^{\dagger}(t)\tilde{G}_{12}(t)(\zeta/\Delta)(\hat{K}_- -\tfrac{\Delta}{2})\big\rangle_b\nonumber\\
& + \big\langle \tilde{G}_{21}^{\dagger}(t)\tilde{G}_{11}(t)(\zeta/\Delta)(\hat{K}_+ -\tfrac{\Delta}{2})\big\rangle_b
\label{rotated_spin_pol}
\end{align}
In the small $\zeta$ limit the second and third terms in \ref{rotated_spin_pol} are negligible, being at least of order $\zeta^2$. For the final two terms, one can estimate the relative size of each contribution by using the infinite series representation of the Greens functions in Eq. \ref{transition_greens_inf_series_non_local} to yield
\begin{align}
&\big\langle \tilde{G}_{11}^{\dagger}(t)\tilde{G}_{12}(t)(\zeta/\Delta)(\hat{K}_- -\tfrac{\Delta}{2})\big\rangle_b =\Bigg\langle\Big(1+i\int_0^t \!\!d\tau_1\hat{\epsilon}(\tau_1)\nonumber\\
&-\int_0^t \!\!d\tau_1 \!\int_0^{\tau_1}\!\!\!d\tau_2  \hat{K}_-(\tau_1)\hat{K}_+(\tau_2)e^{-i\epsilon_{\zeta}(\tau_1-\tau_2)}+...\Big)\nonumber\\
&\times \Big(-i\int_0^t \!\!d\tau_1 \hat{K}_+(\tau_1)e^{-i\epsilon_{\zeta}\tau_1}+...\Big)(\hat{K}_- -\tfrac{\Delta}{2})(\zeta/\Delta) \Bigg\rangle_b \nonumber\\
&\big\langle \tilde{G}_{21}^{\dagger}(t)\tilde{G}_{11}(t)(\zeta/\Delta)(\hat{K}_- -\tfrac{\Delta}{2})\big\rangle_b =\nonumber\\&\Bigg\langle\Big(i\int_0^t \!\!d\tau_1 \hat{K}_+(\tau_1)e^{-i\epsilon_{\zeta}\tau_1}+...\Big)\Big(1-i\int_0^t \!\!d\tau_1\hat{\epsilon}(\tau_1)\nonumber\\
&-\int_0^t \!\!d\tau_1 \!\int_0^{\tau_1}\!\!\!d\tau_2  \hat{K}_+(\tau_1)\hat{K}_-(\tau_2)e^{i\epsilon_{\zeta}(\tau_1-\tau_2)}+...\Big)\nonumber\\
&\times(\hat{K}_+ -\tfrac{\Delta}{2})(\zeta/\Delta)
\Bigg\rangle_b
\end{align}
We next consider the addition of these two correlation functions, as they appear in \ref{rotated_spin_pol} to lowest order in $\zeta$. Averaging over the initial configuration of the bath $\langle K_{\pm}(0)\rangle_b=\langle K\rangle_b$, reveals that the lowest order terms in \ref{rotated_spin_pol} cancel and we are left with
\begin{align}
&\big\langle \tilde{G}_{11}^{\dagger}(t)\tilde{G}_{12}(t)\big\rangle_b(\zeta/\Delta)(\langle K\rangle_b -\tfrac{\Delta}{2})\nonumber\\
& + \big\langle \tilde{G}_{21}^{\dagger}(t)\tilde{G}_{11}(t)\big\rangle_b(\zeta/\Delta)(\langle K\rangle_b -\tfrac{\Delta}{2}) \sim \mathcal{O}\left(\zeta\Delta^3/\omega_c^3 \right)
\end{align}
having applied NIBA to approximate the correlation functions $\langle K_{\pm}(\tau)K_{\mp}(\tau')\rangle\langle K\rangle(\zeta/\Delta)$.
With all these considerations in mind, we find the approximate form for the spin polarization to reduce to
\begin{equation}
\big\langle\sigma_z(t)\big\rangle_b\approx\big\langle \tilde{G}_{11}^{\dagger}(t)\tilde{G}_{11}(t)\big\rangle_b 
\end{equation}

\section{Nearest-Neighbor Blip interactions in the SB model}
\label{NNBA_appendix}

When taking the thermal average over boson operators in Eq. \ref{rho_expanded}, one has correlations between boson clouds (\ref{disp_op}) to all orders. NIBA assumes only `nearest-neighbour' interactions between boson clouds are retained or alternatively, as the name suggests, the `blips' which include pairs of boson cloud operators, are considered noninteracting. Here, we would like to outline our procedure for including 'next nearest-neighbor' boson cloud interactions a.k.a the nearest-neighbour blip approximation (NN), as it is more commonly known N.b. a `blip' is equivalent to two consecutive boson clouds, hence the two different distinctions of NN. 

We will apply NN to the SB model as a point of comparison with our DC model. Therefore, we simply state here the infinite series expression for the dynamics within NIBA
\begin{align}
&\langle\sigma_z(t)\rangle_{b}= 1-\int_0^t \!\!d\tau \int_0^{\tau} \!\!d\tau' \,\,\Sigma_{\text{SB}}(\tau-\tau')\sigma_z(\tau')
\label{inf_series_SB} 
\end{align}
where
\begin{gather}
\Sigma_{\text{SB}}(\tau-\tau') = e^{i\epsilon(\tau-\tau')}\Big\langle B_+(\tau)B_-(\tau') \Big\rangle_b \nonumber\\
+ e^{-i\epsilon(\tau-\tau')}\Big\langle B_-(\tau)B_+(\tau') \Big\rangle_b
\label{}
\end{gather}
which can be found by setting $\zeta=0$ if starting from the DC model.  

Based on methods outlined in,\cite{esquinazi98, weiss08} one can calculate a correction self-energy $\Sigma'_{\text{SB}}$ corresponding to a partial resummation of the series expansion of Eq. \ref{inf_series_SB}. This method amounts to including correlations between adjacent pairs of $B_{\pm}$ operators belonging to different factors of $\Sigma_{\text{SB}}$, in addition to nearest-neighbour correlations already included.

The resulting dynamics, including this correction, is found to be
\begin{equation}
\langle\sigma_z(t)\rangle_{b} ^{\text{NN}} = 1-\int_0^t \!\!d\tau \int_0^{\tau} \!\!d\tau' \,\,\Sigma_{\text{NN}}(\tau-\tau')\sigma_z(\tau')
\end{equation}
where $\Sigma_{\text{NN}} = \Sigma_{\text{SB}}+ \Sigma_{\text{SB}}'$ (NN denotes the nearest-neighbour blip approximation as applied to the SB model). The series can be summed by taking its Laplace transform such that  $\langle\sigma_z(\lambda)\rangle_{b}^{\text{NN}} =(\lambda+\Sigma_{\text{NN}}(\lambda))^{-1}$ where the self-energy correction is calculated to be
\begin{equation}
\Sigma_{\text{SB}}' = \frac{\tilde{\Delta}^2}{\lambda + \tilde{\Delta}^2\text{cosh}(\varphi(t))}
\end{equation}

\end{document}